\documentclass[aps,prb,twocolumn,showpacs,floatfix,superscriptaddress]{revtex4}
\usepackage{dcolumn}
\usepackage{bm}
\usepackage{amsmath,amssymb,graphicx}

\begin{document}
\preprint{Submitted to Phys. Rev. B}

\title{Collective Antenna Effects in the Terahertz and Infrared Response of \\Highly Aligned Carbon Nanotube Arrays}

\author{L.~Ren}
\affiliation{Department of Electrical and Computer Engineering, Rice University, Houston, Texas 77005, USA}

\author{Q.~Zhang}
\affiliation{Department of Electrical and Computer Engineering, Rice University, Houston, Texas 77005, USA}

\author{C.~L.~Pint}
\affiliation{Department of Chemistry, Rice University, Houston, Texas 77005, USA}
\affiliation{Department of Mechanical Engineering, Vanderbilt University, Nashville, Tennessee 37240, USA}

\author{A.~K.~W\'{o}jcik}
\affiliation{Department of Physics and Astronomy, Texas A\&M University, College Station, Texas 77843, USA}

\author{M.~Bunney, Jr.}
\affiliation{NanoJapan Program, Department of Electrical and Computer Engineering, Rice University, Houston, Texas 77005, USA}
\affiliation{Department of Electrical Engineering, Cornell University, Ithaca, New York 14850, USA}

\author{T.~Arikawa}
\affiliation{Department of Electrical and Computer Engineering, Rice University, Houston, Texas 77005, USA}

\author{I.~Kawayama}
\affiliation{Institute of Laser Engineering, Osaka University, Yamadaoka 2-6, Suita, Osaka 565-0871, Japan}

\author{M.~Tonouchi}
\affiliation{Institute of Laser Engineering, Osaka University, Yamadaoka 2-6, Suita, Osaka 565-0871, Japan}

\author{R.~H.~Hauge}
\affiliation{Department of Chemistry, Rice University, Houston, Texas 77005, USA}

\author{A.~A.~Belyanin}
\affiliation{Department of Physics and Astronomy, Texas A\&M University, College Station, Texas 77843, USA}

\author{J.~Kono}
\email[]{kono@rice.edu}
\thanks{corresponding author.}
\affiliation{Department of Electrical and Computer Engineering, Rice University, Houston, Texas 77005, USA}
\affiliation{Department of Physics and Astronomy, Rice University, Houston, Texas 77005, USA}

\date{\today}
\begin{abstract}
We study macroscopically-aligned single-wall carbon nanotube arrays with uniform lengths via polarization-dependent terahertz and infrared transmission spectroscopy.  Polarization anisotropy is extreme at frequencies less than $\sim$3\,THz with no sign of attenuation when the polarization is perpendicular to the alignment direction.  The attenuation for both parallel and perpendicular polarizations increases with increasing frequency, exhibiting a pronounced and broad peak around 10\,THz in the parallel case.  We model the electromagnetic response of the sample by taking into account both radiative scattering and absorption losses.  We show that our sample acts as an effective antenna due to the high degree of alignment, exhibiting much larger radiative scattering than absorption in the mid/far-infrared range.  Our calculated attenuation spectrum clearly shows a non-Drude peak at $\sim$10\,THz in agreement with the experiment.
\end{abstract}

\maketitle


Carbon nanomaterials --- single-wall carbon nanotubes (SWCNTs) and graphene --- have recently emerged as novel terahertz (THz) systems, offering new opportunities for basic research and device applications in THz science and technology.\cite{PortnoietAl06SPIE,Mikhailov09MJ,RenetAl12JIMT,OtsujietAl12JPD}  A variety of proposals exist for using carbon nanotubes for THz devices, including THz sources based on ballistic quasi-metallic SWCNTs\cite{KibisetAl07NL,PortnoietAl08SM} and THz nanoantennas utilizing plasmons in armchair SWCNTs.\cite{Hanson05IEEE,SlepyanetAl06PRB,HaoHanson06IEEE,NemilentsauetAl07PRL}   THz and infrared spectroscopy experiments on SWCNTs reported during the past decade\cite{BommelietAl96SSC,UgawaetAl99PRB,ItkisetAl02NL,JeonetAl02APL,JeonetAl04JAP,JeonetAl05JAP,AkimaetAl06AM,BorondicsetAl06PRB,NishimuraetAl07APL,KampfrathetAl08PRL,RenetAl09NL} have produced conflicting results as well as contradicting interpretations, including the controversial origin of the absorption peak observed around 4\,THz.\cite{UgawaetAl99PRB,AkimaetAl06AM,BorondicsetAl06PRB,NishimuraetAl07APL,KampfrathetAl08PRL,NakanishiAndo09JPSJ,SlepyanetAl10PRB,HartmannetAl11PRB}  Unfortunately, there have been considerable uncertainties and shortcomings with the samples used in the previous studies: not only they were grown by various methods and placed in a variety of THz-transparent polymer films, but also most samples consisted of randomly-oriented bundles of both semiconducting and metallic nanotubes with a wide distribution of lengths and diameters.

Jeon {\it et al.}\cite{JeonetAl02APL,JeonetAl04JAP}~and Akima {\it et al.}\cite{AkimaetAl06AM}~used mechanically stretched nanotube samples with partial tube alignment, observing some polarization anisotropy in THz transmission experiments.  More recently, using extremely well aligned and ultralong SWCNTs, Ren and co-workers demonstrated that carbon nanotubes can act as nearly perfect polarizers in the THz range.\cite{RenetAl09NL}  Kyoung and co-workers, using well-aligned multiwall carbon nanotubes, also demonstrated similarly strong THz polarization anisotropy.\cite{KyoungetAl11NL}  Ren {\it et al.}\cite{RenetAl12NL} has increased the effective thickness of their polarizers to improve the performance of their THz polarizers, demonstrating ideal broadband THz properties: 99.9\% degree of polarization and extinction ratios of 10$^{-3}$ (or 30\,dB) from $\sim$0.4 to 2.2\,THz.

Here we present results of our experimental and theoretical study of the THz and infrared response of highly aligned SWCNT films, similar to the samples used in Refs.~\onlinecite{RenetAl09NL} and \onlinecite{RenetAl12NL}, to elucidate the frequency and polarization dependence of transmission.  Specifically, in the low-frequency THz regime ($<$3\,THz), there is virtually no attenuation when the THz polarization is perpendicular to the nanotube axis but strong attenuation when parallel.  The attenuation for both parallel and perpendicular polarizations increases with increasing frequency, exhibiting a pronounced and broad peak around 10\,THz in the parallel case.  Our theory takes into account both the scattering (or radiative) and absorption (or Ohmic) contributions to the total attenuation of THz and infrared waves through the film.  Although an individual nanotube is an inefficient radiator due to its small diameter, at long wavelengths a large number of aligned nanotubes in the film can be excited coherently and radiate in phase.  The total radiated power is then proportional to the square of the number of nanotubes, $N$, within the coherence area and can be larger than the absorptive loss, which is proportional to $N$ for any frequency.  The total normalized attenuation clearly shows a non-Drude peak at $\sim$10\,THz in agreement with the experiment, indicating that the transmission spectra in the THz and mid-infrared range are primarily determined by the scattering (radiative) loss.

\begin{figure}
\includegraphics[scale=0.47]{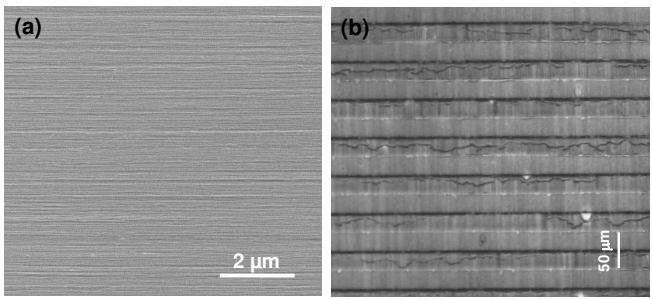}
\caption{\label{spl-images}(color online) (a)~Scanning electron microscope image of a SWCNT film on a sapphire substrate showing the high degree of nanotube alignment. (b)~Optical microscope image of the SWCNT film showing the high uniformity of the nanotube lengths.}
\end{figure}

The highly-aligned SWCNT samples used in this study were synthesized via chemical vapor deposition, as described in Refs.~\onlinecite{XuetAl06JACS,PintetAl08NL,PintetAl08JNN,PintetAl08ACS,PintetAl10ACS}. The as-grown lines of aligned SWCNTs are vertically oriented with respect to the growth substrate, their length determined by the duration of catalyst exposure to the growth conditions. A high temperature (750$^{\circ}$C) H$_{2}$O vapor etch was then used to free the catalyst-SWCNT interface,\cite{PintetAl08ACS} allowing the resulting aligned film to be efficiently transferred to a host substrate ($c$-plane sapphire or undoped silicon in our case).  
The transfer process results in a homogeneous film (initially $\sim$2~$\mu$m thick) that remains as-grown, highly-aligned, and free of exposure to any solvent or liquid.  Figure~\ref{spl-images}(a) shows a top-down scanning electron microscope image of the SWCNTs showing alignment present in such a transferred film. 
Figure~\ref{spl-images}(b) shows a top-down optical microscope image of this horizontally aligned SWCNT film, indicating the high uniformity of the nanotube lengths ($\sim$75\,$\mu$m in this case).

\begin{figure}
\includegraphics[scale=0.56]{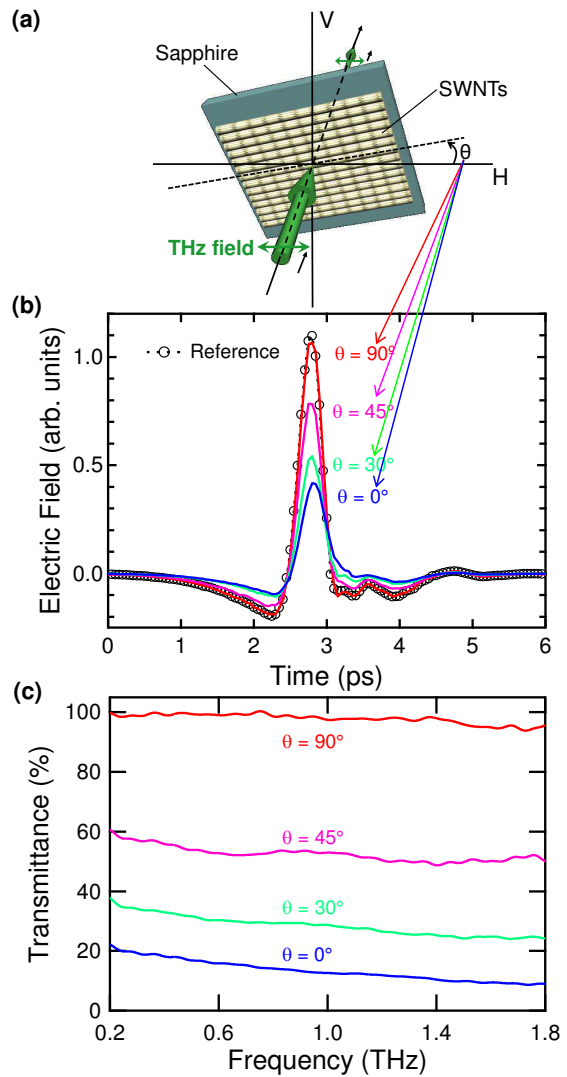}
\caption{\label{THz-data}(color online) (a)~Sketch of experimental configuration, showing the interaction between the linearly polarized THz electric field and the highly aligned SWCNT film. The angle between the THz polarization direction and the nanotube alignment direction, $\theta$, was tuned between 0$^{\circ}$ and 90$^{\circ}$. (b)~Transmitted THz waveforms in the time domain for the reference sapphire substrate (black dashed circled curve) and for the SWCNT film for different angles (colored solid curves) between the THz polarization direction and the nanotube alignment direction. (c)~THz transmittance spectra for four different polarization angles. As the angle between the THz polarization and nanotube alignment direction increases, the transmittance monotonically increases.}
\end{figure}

We used time-domain THz spectroscopy (TDTS) in the 7--100\,cm$^{-1}$ range and Fourier-transform infrared spectroscopy (FTIR) in the 100--7000\,cm$^{-1}$ range.  The TDTS setup used was a typical system based on photoconductive antennas made from low-temperature grown GaAs.\cite{WangetAl07OL,RenetAl09NL}  The THz beam from the emitter was already highly linearly polarized, but a free standing wire-grid polarizer was placed eight inches from both the emitter and the sample to increase
the polarization degree of the incident THz beam.
As schematically shown in Fig.~\ref{THz-data}(a), the SWCNT sample was rotated about the propagation direction of the THz wave. This angle, $\theta$, between the nanotube axis and the THz polarization direction was varied from 0$^{\circ}$ to 90$^{\circ}$.  Polarization-dependent THz transmission measurements were performed on both the SWCNT film sample on a sapphire substrate and a reference sapphire sample with the same thickness as the sample substrate.

Figure~\ref{THz-data}(b) shows the transmitted time-domain waveforms for four different angles ($\theta$ = 0$^{\circ}$, 30$^{\circ}$, 45$^{\circ}$, and 90$^{\circ}$), along with the transmitted waveform from the reference sapphire substrate.  The 90-degree trace is particularly of note. It closely follows the reference waveform, meaning that there is no attenuation when the polarization of the incident beam is perpendicular to the sample's nanotube alignment.  When $\theta$ = 0$^{\circ}$ (parallel polarization), however, strong attenuation of the THz wave is seen. The amount of THz attentuation is more evident in the frequency domain after Fourier-transforming the time-domain waveforms and calculating transmittance spectra, as shown in Fig.~\ref{THz-data}(c).  Here, we plot the transmittance, $T = |\tilde{E_s}/\tilde{E_r}|^2$, as a function of frequency in the 0.2-1.8 THz range, where $\tilde{E_s}$ and $\tilde{E_r}$ are the complex THz signals in the frequency domain for the sample (SWCNT film on sapphire substrate) and reference (sapphire substrate only), respectively.  From this figure, we see that the transmittance of the SWCNT film increases monotonically as the angle $\theta$ increases from 0$^{\circ}$ to 90$^{\circ}$.  When $\theta$ = 90$^{\circ}$, the transmittance is one in this frequency range. Conversely, when $\theta$ = 0$^{\circ}$, the absorbance is finite and the transmittance is much lower, decreasing with increasing frequency to a value lower than 10\% at 1.8~THz.  The 30$^{\circ}$ and 45$^{\circ}$ transmittance lines show the same trend as the 0$^{\circ}$ curve but with larger amplitudes.

\begin{figure}
\includegraphics[scale=0.7]{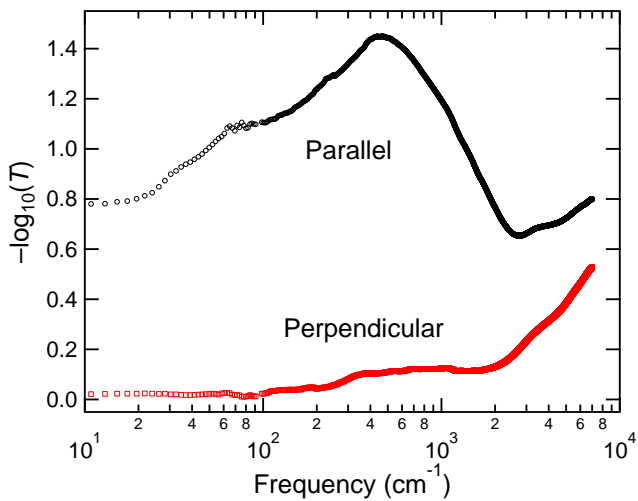}
\caption{\label{FTIR} THz and infrared attenuation spectra for a highly aligned SWCNT film for polarization parallel (black) and perpendicular (red) to the nanotube alignment direction.}
\end{figure}

Figure~\ref{FTIR} shows the attenuation, $-\log_{10}(T)$, for $\theta$ = 0$^{\circ}$ (parallel polarization) to 90$^{\circ}$ (perpendicular polarization) in the entire range of 7--7000\,cm$^{-1}$, combining the results obtained with TDTS and FTIR, for an aligned SWCNT film on an undoped silicon substrate.  Again, highly anisotropic behavior is observed for this SWCNT film in this broad frequency region.  When the polarization is parallel with the nanotube alignment direction, there is strong attenuation exhibiting a prominent and broad peak at $\sim$450\,cm$^{-1}$ (or $\sim$13.6\,THz).  When the polarization is perpendicular to the nanotube alignment direction, the attenuation is significantly suppressed, almost zero in the THz range (as noted earlier) and slightly increasing with increasing in the infrared range.  For both parallel and perpendicular cases, the attenuation rapidly increases toward the near-infrared range, above $\sim$3000\,cm$^{-1}$, due to interband transitions.\cite{PintetAl10ACS,BooshehrietAl11PRB}



\begin{figure}
\includegraphics[scale=0.7]{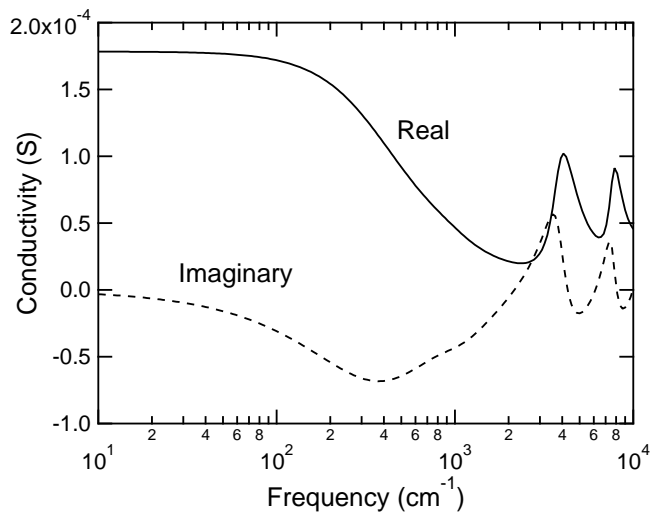}
\caption{\label{fig1_th} Real (solid line) and imaginary (dashed line) parts of the surface conductivity of an infinitely-long metallic single-wall carbon nanotube with a tube diameter of 2.7\,nm, calculated according to Ref.~\onlinecite{SlepyanetAl06PRB}.}
\end{figure}

We model the THz and infrared response of highly aligned SWCNTs by analyzing the electromagnetic scattering and Ohmic losses of a single metallic nanotube.  This is a reasonable approximation since the volume filling factor by nanotubes is only a few percent; also, we are mainly interested in modeling the long-wavelength response, for which the contribution of semiconducting nanotubes is negligible.  The surface conductivity of the nanotubes, $\sigma_{\rm c}$, is calculated numerically, using the model of rolled-up graphene (see, e.g., Refs.~\onlinecite{SlepyanetAl06PRB} and \onlinecite{HaoHanson06IEEE}). We assume the phenomenological relaxation time, $\tau$, to be equal to 10\,fs in the THz and mid-infrared range and 5\,fs in the near infrared.  
The real and imaginary parts of the conductivity versus frequency calculated for an infinitely long metallic nanotube with a diameter of 2.7~nm are plotted in Fig.~\ref{fig1_th}.

The electromagnetic response of a nanotube excited by a normally incident plane wave is calculated by solving the modified Hall\'en equation
\begin{eqnarray}
  \int^{L/2}_{-L/2}(K(z-z')+q(z-z'))I(z',\omega)dz' &=&  \nonumber \\ 
  - \frac{i4 \pi \omega \varepsilon_0}{2k} \int_{-L/2}^{L/2} E_0 \sin k|z-z'| dz' \nonumber \\
  + C_1 \sin kz + C_2 \cos kz,
\end{eqnarray}
where $I(z,\omega)$ is the current induced on the nanotube, the expressions for the kernel functions $K$ and $q$ are given in Ref.~\onlinecite{Hanson05IEEE}, $\varepsilon_0$ is the free space permittivity, $k=2 \pi / \lambda$ is the free space wavenumber, $L$ is the nanotube length, $E_0$ is the amplitude of the incident plane wave, and the constants $C_1$ and $C_2$ are obtained from the boundary conditions $I(\pm L/2)=0$.  This integral equation is solved for $I(z,\omega)$ numerically, using the method of moments with a normalized incident wave, $E_0=1$.  The current $I(z,\omega)$ is then used to calculate the scattered (or radiated) field as well as the absorbed field due to Ohmic losses. The total power radiated to the far field into all directions is calculated as
%
\begin{equation}
  P_{\rm r}(L,\omega)= \frac{\omega^2}{4 \epsilon_0 c^3} \int_0^{\pi} d\theta \sin^3 \theta \left| \int_{-L/2}^{L/2} I(z,\omega) dz \right|^2,
\end{equation}
whereas the absorbed power is\cite{WuKing65IEEE}
\begin{equation}
  P_{\rm o}(L,\omega)=\frac{1}{4 \pi R} \textrm{Re}\{\sigma^{-1}_{c}(\omega)\} \int_{-L/2}^{L/2} |I(z,\omega)|^2 dz.
\end{equation}

%

\begin{figure}
\includegraphics[scale=0.75]{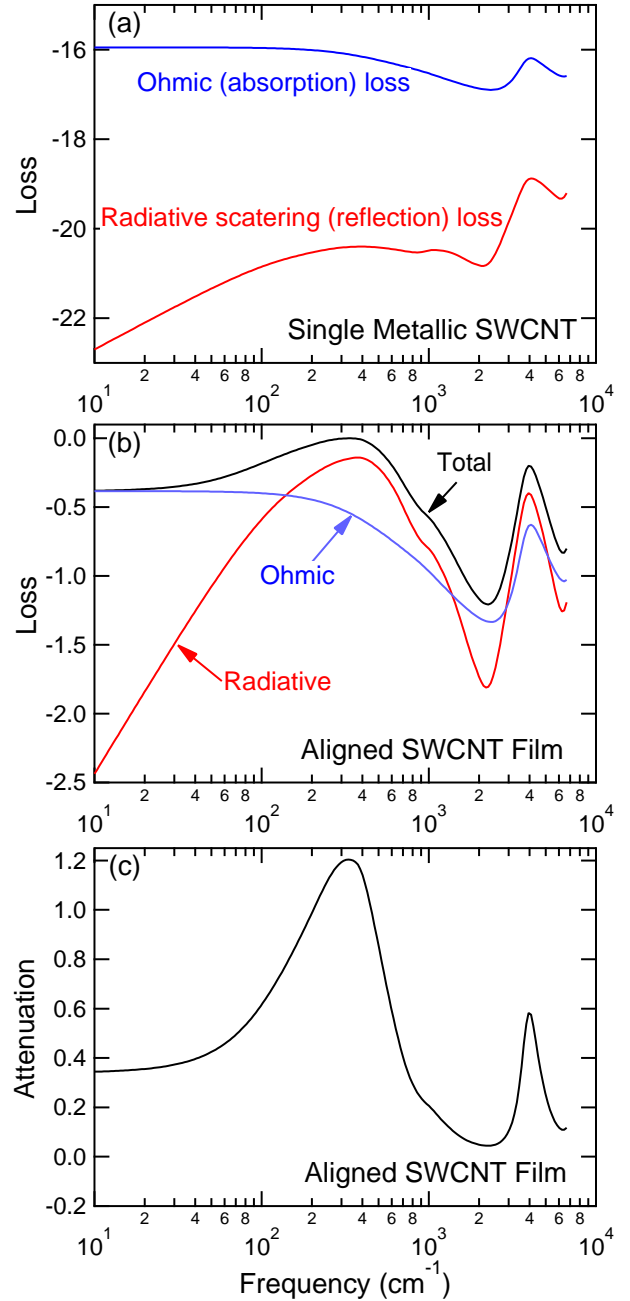}
\caption{\label{fig2_th} (a)~The logarithm of absorptive (top trace) and radiative (bottom trace) loss versus frequency calculated for a single metallic SWCNT with a diameter of 2.7~nm and a length of 75~$\mu$m.  (b)~The logarithm of radiative (red line), Ohmic (blue line), and total (black line) losses for the film, calculated including the collective antenna effect with  $N=5\times 10^4$ and assuming metallic SWCNTs of the same diameter of 2.7\,nm and a length of 75\,$\mu$m.   (c)~Total normalized attenuation of the film, calculated from the total losses shown in (b).} 
\end{figure}

The total attenuation of the incident electromagnetic wave by a film is calculated by adding $P_{\rm r}$ and $P_{\rm o}$.  A single nanotube is an inefficient radiator due to its small radius,\cite{Hanson05IEEE} as also shown in Fig.~\ref{fig2_th}(a). However, at long wavelengths a large number of highly aligned nanotubes in the film can be excited coherently and radiates in phase.  The total radiated power $P_{\rm r}$  is then proportional to the {\em square} of the number, $N$, of nanotubes within the coherence area which scales roughly as $\lambda^{2}$.  As a result, the radiated power can become larger than the absorbed power since the latter $P_{\rm o} \propto N$ for any frequency.\cite{ShubaetAl07PRB}  We assume an $N^2$ dependence for the radiated power at long wavelengths up to about 20\,$\mu$m, followed by a $\lambda^{2}$ cutoff to the $\propto N$ dependence.

The spectrum of Ohmic (absorption) and radiative scattering (reflection) losses for a single metallic nanotube with a diameter of 2.7~nm and a length of 75~$\mu$m is presented in Fig.~\ref{fig2_th}(a). The spectrum of losses for a film with a total number of $N=5\times 10^4$ nanotubes contributing collectively to scattering at long wavelengths is shown in Fig.~\ref{fig2_th}(b);  the corresponding total attenuation is given in Fig.~\ref{fig2_th}(c). Note the logarithmic scale for losses in Figs.~\ref{fig2_th}(a) and \ref{fig2_th}(b).  Both the loss and attenuation spectra clearly show a non-Drude peak at $\sim$10~THz, in agreement with the experiment. The peak originates from collective radiative scattering.  There is a crossover from a radiation-dominated regime to an absorption-dominated regime at the lowest frequencies below $\sim $2~THz ($\sim$60~cm$^{-1}$) in Fig.~\ref{fig2_th}(b). This happens because radiative losses drop rapidly with decreasing frequency below the peak whereas the absorption losses increase and approach a constant value. This crossover explains the behavior of the measured attenuation spectrum in the lowest frequency range in Fig.~\ref{FTIR}, which approaches a plateau. The fact that absorption becomes important at the longest wavelengths is further supported by direct calculations of absorbance based on the complex refractive index in the 0.2-2~THz range extracted from our TDTS experiments: the absorbance of the film is more than 50\% at frequencies below $\sim$0.5~THz.   

The presence of the non-Drude peak at $\sim$10~THz in our calculations does not depend on the position and precise spectral dependence of the cutoff from $N^2$ to $N$ dependence in the radiative losses of the film.  The only important factor is that the attenuation in the mid/far-infrared wavelength range should be mostly due to collective radiative scattering, not absorptive Ohmic losses; otherwise, the spectrum of these long nanotubes ($L \gg 1$~$\mu$m) should be Drude-like at these wavelengths.\cite{NakanishiAndo09JPSJ}  Including a spread of nanotube lengths and diameters would make the peak broader and may affect its shape/asymmetry.  The second peak in the near-infrared ($>$3000~cm$^{-1}$) is due to interband transitions.


In summary, polarization dependent THz time-domain spectroscopy and Fourier-transform infrared spectroscopy of highly-aligned single-wall carbon nanotube films revealed strongly anisotropic response in the entire spectral range.  In the low-frequency THz regime ($<$3~THz), there is virtually no absorption when the THz polarization is perpendicular to the nanotube axis.  The attenuation for both parallel and perpendicular polarizations increases with increasing frequency, exhibiting a pronounced and broad peak around 10~THz in the parallel case.  Our theory, which takes into account both coherent radiative scattering and absorptive Ohmic losses, successfully reproduces the observed spectra.  The calculated attenuation spectrum clearly shows a non-Drude peak at $\sim$10~THz, which indicates that the transmission spectra in the THz and infrared range are primarily determined by radiative scattering losses, reflecting the collective antenna properties of these aligned, ultralong carbon nanotubes.



This work was supported by the Department of Energy (through Grant No.~DE-FG02-06ER46308), the National Science Foundation (through Grants No.~OISE-0530220 and EEC-0540832), and the Robert A.~Welch Foundation (through Grant No. C-1509).  




\end{document}